\documentclass[usegraphicx,usenatbib,useapjfonts,apj]{emulateapj}
\begin{document}
\newcommand{\vk}{{\bf k}}

\def\plotone#1{\centering \leavevmode
\epsfxsize=\columnwidth \epsfbox{#1}}

\title[The effect of primordial non-Gaussianity on halo bias]{The effect of primordial non-Gaussianity 
on halo bias}

\author{Sabino Matarrese\altaffilmark{1} \&  Licia Verde \altaffilmark{2,3}}
\altaffiltext{1}{Dipartimento di Fisica "G. Galilei", Universit\`a 
degli Studi di Padova and INFN, Sezione di Padova, via Marzolo 8, 35131, Padova, Italy; sabino.matarrese@pd.infn.it}
\altaffiltext{2}{ICREA and Institute of Space Sciences (CSIC-IEEC), UAB, Barcelona 08193, Spain}
\altaffiltext{3}{Department of Astrophysical Sciences, Peyton Hall, Princeton University, Ivy Lane, NJ 08544, USA ; verde@ieec.uab.es}

\begin{abstract}

It has long been known how to  analytically relate the clustering properties of the collapsed 
structures (halos)  to those of the underlying dark matter distribution for Gaussian initial conditions. 
Here we apply the same approach to physically motivated non-Gaussian models.
The techniques we use were developed in the 1980s to deal with the clustering of 
peaks of non-Gaussian density fields. 
The description of the clustering of halos for non-Gaussian initial conditions has  recently received renewed interest, motivated by the forthcoming  large galaxy and cluster surveys. For inflationary-motivated non-Gaussianites, we find an analytic expression for  the halo bias  as a function of scale, mass and redshift, employing only the approximations of high-peaks and large separations. 

\end{abstract}


\section{Introduction}
Constraining primordial non-Gaussianity offers a powerful test  of the generation mechanism of cosmological perturbations in the early universe. While standard single-field models of slow-roll inflation lead to small departures from Gaussianity, non-standard scenarios  allow for a larger level of non-Gaussianity (\citet{BKMR04} and references therein). The standard observables to constrain non-Gaussianity are the cosmic microwave background and large-scale structure. A powerful technique is based on the abundance \citep{MVJ00, VJKM01, Loverdeetal07, RB00, RGS00} and clustering \citep{GW86, MLB86, LMV88} of rare events such as  dark matter density peaks  as they trace the tail of the underlying distribution.  These theoretical predictions have been tested against numerical N-body simulations \citep{KNS07,Grossietal07,DDHS07}.  \citet{DDHS07} showed that primordial non-Gaussianity affects the clustering of dark matter halos inducing a scale-dependent bias.  This effect will be useful for constraining non-Gaussianity from  future surveys which will provide a large sample of galaxy clusters over a volume comparable to the horizon size (e.g., DES, PanSTARRS, PAU, LSST, DUNE, ADEPT, SPACE, DUO) or mass-selected  large clusters  samples via the Sunyaev-Zel'dovich effect (e.g., ACT, SPT),  considered alone or via cross-correlation techniques (e.g., ISW, lensing).

Here, we resort to results and techniques developed in the 1980s \citep{GW86, MLB86, LMV88} to  extend this work and derive an accurate analytical expression for halo bias,  in the presence of general non-Gaussian initial conditions, accounting for  its scale, mass and redshift dependence.

\section{Halos as peaks of the density field}

Halo clustering is generally studied by assuming that halos correspond to regions 
where the (smoothed) linear density field exceeds a suitable threshold. This 
amounts to modeling the local halo number density as a theta (step) function 
\begin{equation}
\rho_{\rm h, R}({\bf x}, z_f) \!= \!
\theta\left[\delta_R({\bf x},z_f)\!-\!\Delta_c \right]\! =\!
\theta\left[\delta_R({\bf x})\!-\!\delta_c(z_f) \right], 
\end{equation}
modulo a constant normalization factor which is irrelevant 
for the calculation of correlations. Here $R$ denotes a smoothing radius which defines  
the halo mass $M$ by $M=\Omega_{m,0}3 H_0^2/(8 \pi G) (4/3) \pi R^3$, with $\Omega_{m,0}$ denoting the present-day matter density parameter, $H_0$ the present-day Hubble parameter and $G$ Newton's constant. The threshold $\Delta_c$ is the linearly extrapolated over-density 
for spherical collapse: it is $1.686$ in the Einstein-de Sitter case, 
while it slightly depends on redshift for more general 
cosmologies (e.g., \citet{KS96}). 
The redshift $z_f$ is the formation redshift of the halo, 
which for high mass halos is very close to the observed redshift $z_o$. 
Hereafter we will thus make the approximation $z_f\simeq z_o=z$. The second equality can be understood if  we think of the density fluctuation as 
being time-independent while giving a redshift dependence to the collapse threshold 
$\delta_c (z_f ) \equiv  \Delta_c (z_f )/D(z_f)$. Here 
$D(z)$ denotes the general expression for the linear growth factor, 
which depends on the background cosmology. In particular $D(z)=(1+z)^{-1}g(z)/g(0)$ where $g(z)$ is the growth suppression factor  for non Einstein-de Sitter Universes.

For Gaussian initial conditions, 
one obtains (\cite{Kaiser84, PolitzerWise84, JensenSzalay86})  
\begin{equation}
\xi_{h,M}(r)=\exp\left[ \frac{\nu^2}{\sigma_R^2}\xi_R(r)\right]-1
\simeq \frac{\nu^2}{\sigma_R^2}\xi_R(r) \;, 
\label{eq:exponent}
\end{equation}
where   $\sigma_R$ is  the {\it r.m.s.} 
of the underlying dark matter fluctuation field smoothed on scale $R$,  $\nu=\delta_c/\sigma_R$ and $\xi_{h,M(R)}$ denotes the correlation function 
of the halos of mass $M$ corresponding to radius $R$. $W_R$ denotes the 
top-hat function of width $R$ and the definition of $\xi_R(r)$ 
is $\int d^3r'\xi(r')W^2_R(|{\bf r}-{\bf r}'|)$. 
In the second equality above we have expanded the exponential in series. 
The truncation of the series holds for separations $r \gg R$.
Thus we obtain the well-known Kaiser's formula \citep{Kaiser84} of a 
scale-independent bias:
\begin{equation}
\xi_{h,M}(r)=b_L^2 \xi_R(r)
\end{equation}
where $b_L=\delta_c/\sigma_R^2$. Compared with the more refined relations in 
e.g., \citet{MoWhite96} and  \citet{CLMP98}, an additive term 
$1/\delta_c$ has been dropped as a consequence of the high-peak 
approximation in the first equality of Eq.~(\ref{eq:exponent}).

Here the subscript $L$ indicates that this should be considered as a 
Lagrangian bias, because all correlations and peaks considered here 
are those of the {\it initial} density field (linearly extrapolated 
till the present time). 
Making the standard assumptions that halos move coherently 
with the underlying dark matter, one can obtain the 
final Eulerian bias as $b_E=1+b_L$, using the 
techniques outlined in  \citet{Eetal88}, \citet{CK89}, \citet{MoWhite96} and  \citet{CLMP98}.

The two-point correlation function of regions above a high threshold has 
been obtained,  for the general non-Gaussian case, in \citet{GW86}, \citet{MLB86} and \citet{LMV88}:
\begin{eqnarray}
\xi_{h, M}(|{\bf x}_1-{\bf x}_2|)=-1+\;\;\;\;\;\;\;\;\;\;\;\;\;\;\;\;\;\;\;\;\;\;\;\;\;\;\;\;\;\;\\ 
\exp\left\{\sum_{N=2}^{\infty}
\sum_{j=1}^{N-1}\frac{\nu^N\sigma_R^{-N}}{j!(N-1)!}\xi^{(N)}
\left[^{{\bf x}_1,...,{\bf x}_1, \,\,\,{\bf x}_2,........, {\bf x}_2}_{j \,times  
\,\,\,\,\,\, (N-j)\, times}\right]\right\} \;. \nonumber
\end{eqnarray} 

As before, for large separations we can expand the exponential to first order.
To leading order for non-Gaussianity of the type \citep{SalopekBond90, Ganguietal94,VWHK00, KS01}
\begin{equation}
\Phi=\phi+f_{\rm NL}*\left(\phi^2-\langle \phi^2 \rangle\right)
\label{eq:fnl}
\end{equation}
where $*$ denotes convolution, as in general $f_{\rm NL}$ may be scale and configuration 
dependent, but for constant $f_{\rm NL}$ it reduces to a simple multiplication. 
For simplicity, below we will carry out calculations assuming constant $f_{\rm NL}$ and 
will generalize our results at the end. 
Here $\Phi$ denotes Bardeen's gauge-invariant potential, which, 
on sub-Hubble scales reduces to the usual Newtonian peculiar 
gravitational potential, up to a minus sign.  In the literature, there are two conventions for Eq.~(\ref{eq:fnl}): the large-scale structure and the CMB one. Following the large-scale structure convention, here $\Phi$ is linearly extrapolated at $z=0$. In the CMB convention $\Phi$ is instead primordial:  thus $f_{\rm NL}=g(z=\infty)/g(0) f_{\rm NL}^{CMB}$. 
In Eq.~(\ref{eq:fnl}), $\phi$ denotes a Gaussian random field. 
For values of $f_{\rm NL}$ consistent with  observations, 
we can keep terms up to the three-point correlation function $\xi^{(3)}$, 
obtaining that the correction to the halo correlation function, $\Delta \xi_{h}$ 
due to a non-zero three-point function is given by:
\begin{eqnarray}
\label{eq:correxi}
\Delta \xi_{h}&=&\frac{\nu_R^3}{2\sigma_R^3}
\left[\xi_R^{(3)}({\bf x}_1,{\bf x}_2,{\bf x}_2)+\xi_R^{(3)}({\bf x}_1,{\bf x}_1,{\bf x}_2)
\right]  \nonumber \\ 
&=& \frac{\nu_R^3}{\sigma_R^3}\xi_R^{(3)}({\bf x}_1,{\bf x}_1,{\bf x}_2)
\end{eqnarray}
 
\section{Application to a local non-Gaussian model}

We want to find  an expression for the correlation function 
of the late-time halos which form from the dark matter over-density.
  
In Fourier space, the present-time ($z=0$) filtered linear over-density 
$\delta_R$ is related to $\Phi$ by the Poisson equation:
\begin{equation}
\delta_{R}({\bf k})=\frac{2}{3} \frac{T(k) k^2}{H_0^2 
\Omega_{m,0}}W_R(k)\Phi({\bf k})\equiv {\cal M}_R(k)\Phi({\bf k}) \;,
\label{eq:defM}
\end{equation}
where $T(k)$ denotes the matter transfer function\footnote{The matter transfer function  and the window functions cannot be neglected here: the initial conditions of Eq.~(\ref{eq:fnl}) are set out 
well before matter-radiation equality and the density field should be smoothed  to define the halo mass.} and $W_R(k)$ is the Fourier transform of  $W_R({\bf r})$.
From Eq.~(\ref{eq:fnl}) the definition of $\Phi$ is 

\begin{equation}
\Phi({\bf k})=\phi({\bf k})+f_{\rm NL}\int \frac{d^3k'}{(2\pi)^3}
\phi({\bf k}')\phi({\bf k}-{\bf k}') \;,
\end{equation}
whose bispectrum is 
\begin{eqnarray}
B_{\phi}(k_1,k_2,k_3)&=&2 f_{\rm NL}\left[ P_{\phi} (k_1) P_{\phi}(k_2)\right. \nonumber \\ 
&+&\left.P_{\phi} (k_2) 
P_{\phi}(k_3)+P_{\phi} (k_1) P_{\phi}(k_3) \right]\;.
\end{eqnarray}

With these definitions the density bispectrum becomes
$ B_{\delta}(k_1,k_2,k_3)={\cal M}_R(k_1)
{\cal M}_R(k_2){\cal M}_R(k_3)B_{\phi}(k_1,k_2,k_3)$,
where $P_{\phi}$ denotes the power-spectrum of the Gaussian field $\phi$. 
The three-point function of Eq.~(\ref{eq:correxi}) becomes
\begin{eqnarray}
\label{eq:xi3}
&\xi^{(3)}({\bf x}_1,{\bf x}_1,{\bf x}_2)=&   \\ 
&\frac{1}{(2 \pi)^6}\int d^3k_1 d^3k_2 2f_{\rm NL} 
{\cal M}_R(k_1){\cal M}_R(k_2){\cal M}_R(|\vk_1+\vk_2|)\times & \nonumber \\
& \left[P_{\phi}(k_1)P_{\phi}(k_2)+2P_{\phi}(k_1)
P_{\phi}(|\vk_1+\vk_2|)\right] e^{i (\vk_1+\vk_2)\cdot (x_1-x_2)}& \nonumber
\end{eqnarray} 
%
%
and  Fourier transform  of Eq.~(\ref{eq:xi3}) becomes
\begin{eqnarray}
\frac{2f_{\rm NL}}{(2\pi)^2}{\cal M}_R(k)\int dk_1 k_1^2 {\cal M}_R(k_1)
P_{\phi}(k_1) \times \nonumber \\
\!\int_{-1}^1\!d\mu
{\cal M}_R\left(\sqrt{\alpha}\right)\left[ \!P_{\phi}\left(\sqrt{\alpha}\right) 
+ 2 P_{\phi}(k) \right]  \nonumber
\end{eqnarray}
where $\alpha=k_1^2+k^2+2k_1k\mu$.

\section{Results}
\begin{figure}
\plotone{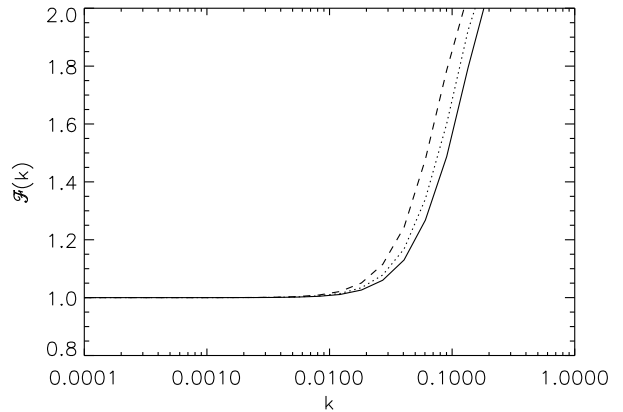}
\caption{The function  ${\cal F}_R(k)$ for three different masses: $1\times 10^{14}$ M$_{\odot}$ (solid), $2\times 10^{14}$ M$_{\odot}$ (dotted), $1\times 10^{15}$ M$_{\odot}$ (dashed).}
\label{fig:FR}
\end{figure}

We can now write an expression for the non-Gaussian contribution to the halo power spectrum. 
Eq.~(\ref{eq:correxi}) in Fourier space becomes:
\begin{equation}
\Delta P_{\rm h}(k)= b^2_{0,L} 4 f_{\rm NL} \delta_c P_{\phi \delta}(k) {\cal F}_R(k) 
\label{eq: res1}
\end{equation}
where we have used $b_{0,L} \equiv \delta_c/\sigma_R^2$, corresponding to 
the Lagrangian linear bias that the halos would have in the Gaussian case, 
$P_{\phi \delta}(k) \equiv {\cal M}_R(k)P_{\phi}(k)$ and 
\begin{eqnarray}
{\cal F}_R(k)=\frac{1}{8\pi^2\sigma_R^2}\int dk_1 k_1^2 {\cal M}_R(k_1)
P_{\phi}(k_1) \times \nonumber \\
\!\int_{-1}^1\!d\mu
{\cal M}_R\left(\sqrt{\alpha}\right)\left[ \!\frac{P_{\phi}\left(\sqrt{\alpha}\right)}{P_{\phi}(k)} 
+ 2 \right] . 
\end{eqnarray}
The ``form"  factor ${\cal F}_R(k)$ is plotted  as a function of $k$ in Fig.~\ref{fig:FR} for three different masses.

The expression for the halo power-spectrum can be rewritten in a more convenient form 
where we can also make the redshift dependence explicit:
$$
P_{\rm h}(k,z)=\frac{\delta_c^2(z)P_{\delta \delta}(k, z)}{\sigma_R^4 D^2(z)}
\left[ 1+4 f_{\rm NL}\delta_c(z)\frac{P_{\phi \delta}(k) 
{\cal F}_R(k)}{P_{\delta \delta}(k)}\right]   
$$
where $P_{\delta \delta}(k,z)= D^2(z)P_{\delta \delta}(k)=D^2(z){\cal M}_R^2(k)P_{\phi}(k)$.

\begin{figure}
\plotone{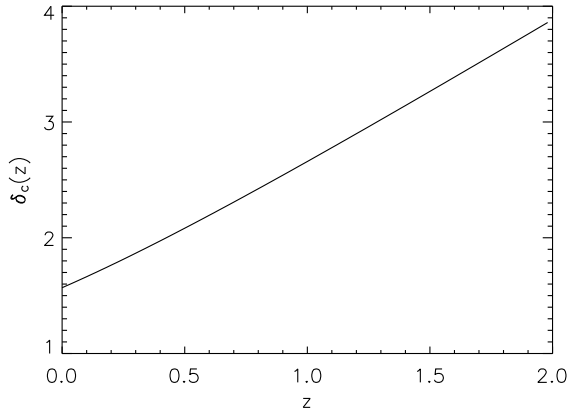}
\caption{The redshift dependence of $\Delta b_h/b_h$.}
\label{fig:z}
\end{figure}

\begin{figure}
\plotone{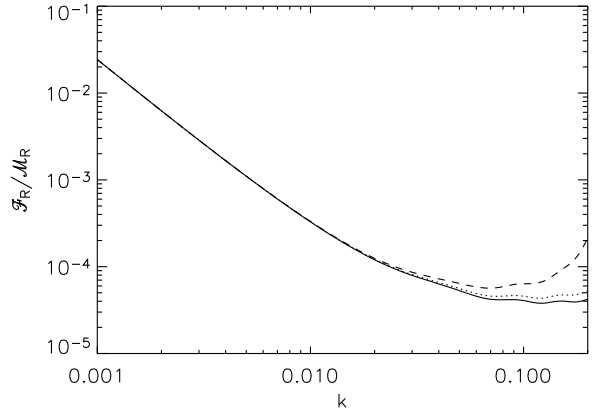}
\caption{The scale dependence of $\Delta b_h/b_h$ for three different masses: $1\times 10^{14}$ M$_{\odot}$ (solid), $2\times 10^{14}$ M$_{\odot}$ (dotted), $1\times 10^{15}$ M$_{\odot}$ (dashed).}
\label{fig:k}
\end{figure}

We can now define the Lagrangian bias $b_L$ of the halos from 
$b^2_L=P_{\rm h}(k,z)/P_{\delta \delta}(k,z)$ and use $b^E=1+b_L$ to obtain the 
expression for the non-Gaussian halo bias
\begin{equation}
b_{\rm h} ^{f_{\rm NL}}= 1+\frac{\Delta_c(z)}{\sigma_R^2 D^2(z)} \left[ 1+2 f_{\rm NL}\frac{\Delta_c(z)}
{D(z)}\frac{{\cal F}_R(k)}{{\cal M}_R(k)} \right] \;.
\end{equation}
Thus $b_h^{f_{\rm NL}}=b_h(1+\Delta b_h/b_h)$ where $b_h$ denotes the halo bias for the Gaussian case. $\Delta b_h/b_h$ is $2 f_{\rm NL}$ times a redshift-dependent factor $\Delta_c(z)/D(z)$, plotted in Fig.~(\ref{fig:z}),  times a $k$ and mass dependent  factor  ${\cal F}_R(k)/{\cal M}_R(k)$, shown in Fig.~\ref{fig:k}.

\section{Discussion and conclusions}
We have obtained an  analytic expression for the bias of dark matter halos for non-Gaussian  initial conditions. 
The only approximations used in our approach are: {\it i)} 
high peaks, {\it i.e.} large values of 
$\nu=\delta_c(z)/\sigma_R$ (as in the original Kaiser's formula), 
which essentially amounts to a limitation on
the mass range over which one can apply this formula, and {\it ii)} 
large separation among the halos, which is the standard assumption
allowing to use linear bias.  While it is true that on large scales ($k\rightarrow 0$)  the form factor, the transfer function and the window function  go to unity, on the scales of interest  neglecting these terms may lead to errors on $\Delta b_h$ and therefore on $f_{\rm NL}$ of the order of 100\%.
Comparison of these analytical findings with simulations  will be presented elsewhere (Grossi {\it et al.}, in preparation).

An advantage of our approach is that it  can be easily generalized to non-local and scale-dependent non-Gaussian models in which  $B_{\phi} (k_1, k_2,k_3)$ is the dominant higher-order correlation and  has a general form, obtaining
\begin{eqnarray}
\frac{\Delta b_h}{b_h}&=& \frac{\Delta_c(z)}{D(z)}\frac{1}{8\pi^2\sigma_R^2}\int dk_1 k_1^2 {\cal M}_R(k_1) \times \nonumber \\
&&\!\int_{-1}^1\!d\mu
{\cal M}_R\left(\sqrt{\alpha}\right)\frac{B_{\phi}(k_1,\sqrt{\alpha},k)}{P_{\phi}(k)}.
\label{eq: fnlk}
\end{eqnarray}
Modeling the clustering of hot and cold CMB spots for non-Gaussian initial conditions, is a straightforward extension of this calculation (Heavens {\it et al.}, in preparation).

We envision that this calculation will be useful for constraining non-Gaussianity from  future surveys which will provide a large sample of galaxy clusters over a volume comparable to the horizon size (e.g., 
DES, PanSTARRS, PAU, LSST, DUNE, ADEPT, SPACE, DUO) or mass-selected (via the Sunyaev-Zel'dovich effect)  large clusters  samples (e.g., ACT, SPT).

\section*{Acknowledgments}
SM acknowledges partial support by ASI contract I/016/07/0 "COFIS".
LV is supported by FP7-PEOPLE-2007-4-3-IRG n. 202182 and CSIC  I3 grant n. 200750I034. LV and SM  thank N. Afshordi for comments.
The authors would like to thank the XIX Canary Islands winter school of Astrophysics, 
where part of this work was carried out and the January 2008 Aspen winter meeting 
at the Aspen Center for Physics where this work was completed.


\end{document}